\documentstyle[12pt,a4]{article}
\newcommand{\drawsquare}[2]{\hbox{%
\rule{#2pt}{#1pt}\hskip-#2pt%
\rule{#1pt}{#2pt}\hskip-#1pt%
\rule[#1pt]{#1pt}{#2pt}}\rule[#1pt]{#2pt}{#2pt}\hskip-#2pt%
\rule{#2pt}{#1pt}}%
\newcommand{\f}{\raisebox{-.5pt}{\drawsquare{6.5}{0.4}}}
\newcommand{\af}{\overline{\f}}

\begin{document}
\baselineskip=16pt
\begin{titlepage}
\begin{flushright}
  KUNS-1739\\[-1mm]
  UT-971\\[-1mm]
  TU-633\\[-1mm]
  hep-ph/0110117
\end{flushright}

\begin{center}
\vspace*{1.5cm}
  
{\Large\bf 4D construction of bulk supersymmetry breaking}
\vspace{15mm}
  
{\large
  Tatsuo~Kobayashi\footnote{E-mail address:
    kobayash@gauge.scphys.kyoto-u.ac.jp},
  Nobuhito~Maru\footnote{E-mail address: 
    maru@hep-th.phys.s.u-tokyo.ac.jp} and
  Koichi~Yoshioka\footnote{E-mail address:
    yoshioka@tuhep.phys.tohoku.ac.jp}
  \vspace{8mm}
  }

$^*${\it Department of Physics, Kyoto University Kyoto 606-8502,
  Japan}\\
$^\dagger${\it Department of Physics, University of Tokyo, Tokyo
  113-0033, Japan}\\
$^\ddagger${\it Department of Physics, Tohoku University, Sendai
  980-8578, Japan}
\vspace{1.6cm}

\begin{abstract}
In this letter, we discuss a four-dimensional model with modulus
fields which are responsible for supersymmetry breaking. Given
non-trivial moduli dependence of the action, the model is found to
give a proper description of higher-dimensional supersymmetry
breaking. We explicitly calculate gaugino and scalar mass spectrum and
show that several classes of scenarios proposed in the literature are
described in certain regions of the parameter space of the moduli
vacuum expectation values. The model in other generic regions of the
moduli space gives unexplored scenarios (mass spectra) of
supersymmetry breaking in four dimensions.
\end{abstract} 
\end{center}
\end{titlepage}


\section{Introduction}
\setcounter{footnote}{0}

Supersymmetry is one of the most interesting ideas which have been
introduced to overcome some unsatisfactory points of the Standard
Model. For example, the gauge coupling unification from the precise
electroweak measurements~\cite{GUT} and the stability of the
Planck/weak mass hierarchy~\cite{hierarchy} are great successes of
phenomenological applications of supersymmetry. It is, however,
experimentally certain that supersymmetry is broken above the weak
scale, while a variety of mechanisms for supersymmetry breaking have
been proposed so far.

Among these, the mechanisms which are involved in higher-dimensional
physics have been extensively studied in various ways. The existence
of extra spatial dimensions provides novel ways to break supersymmetry
and to communicate it to our four-dimensional world, which is the
low-energy effective theory of the models. A well-known framework is
the string-inspired four-dimensional supergravity~\cite{sugra}. In
large classes of these models, there are two modulus fields concerned
with the compactified extra dimensions, called dilaton and overall
modulus, which are assumed to develop non-vanishing vacuum expectation
values (VEV) in their auxiliary components. The supersymmetry-breaking
effect is transmitted to our low-energy degrees of freedom via
(super)gravity interactions. There have been other interesting
mechanisms for supersymmetry breaking with extra
dimensions~\cite{others}.
These approaches provide phenomenologically viable particle spectra 
due to intrinsic nature of higher-dimensional theories. 

In this letter, we present a purely four-dimensional framework which 
can describe supersymmetry breaking in the bulk. To this end, it is
convenient to regard extra dimensions as being
latticized~\cite{ACG,HPW}. With this method, it is possible to revisit
many interesting feature of higher-dimensional effects from the
four-dimensional point of view~\cite{apply}. Thus, models can
incorporate various four-dimensional mechanisms such as for flavor
problems, and at the same time utilize five-dimensional nature stated
above. We study a model with two types of modulus fields which are
supposed to have supersymmetry-breaking VEVs. Given non-trivial moduli
dependences of the action, it is found that certain limits in this
two-dimensional parameter space of VEVs reproduce the mass spectra of
the bulk scenarios in the literature. Other generic regions of the
moduli space give unexplored scenarios for supersymmetry breaking.

In Section 2, we explain our setup and briefly touch on the spectrum of 
vector fields. Supersymmetry breaking (non-zero $F$-terms) in this model 
is discussed in Section 3, where we identify various modulus fields and 
reveal their connections in the light of the construction of model. In 
Section 4, we calculate the mass spectrum of vector multiplets with the 
non-vanishing moduli $F$-terms, and show typical mass splitting in the 
limits that correspond to various supersymmetry-breaking mechanisms in 
higher dimensions. Section 5 is devoted to the summary of our results.

\section{Model}
\setcounter{equation}{0}

We consider a four-dimensional supersymmetric gauge theory with the
$N$ copies of gauge groups $G^N=G_1\times G_2\cdots \times G_N$. We
assume that, for simplicity, all the gauge theories $G_i$'s have the
same structure and particularly have the common gauge coupling
$g$. The $N=1$ vector multiplet $V_i$ of the $G_i$ gauge theory
contains a gauge field $A_\mu^i$ and a gaugino $\lambda^i$. In
addition, there are $N=1$ chiral multiplets $Q_i$ ($\,i=1,\cdots,N$)
in bifundamental representation, that is, $Q_i$ transforms as
$(\f,\,\af)$ under the $(G_i,G_{i+1})$ gauge symmetries. The fields
$Q_i$ are referred to as link variables in that they link up two
neighboring gauge theories. The field content of the theory is
summarized in Table~\ref{model}.
\begin{table}[htbp]
\begin{center}
\begin{tabular}{c|ccccc} 
  & ~$G_1$~ & ~$G_2$~ & ~$G_3$~ & ~$\cdots$~ & ~$G_N$~ \\ \hline
  $Q_1$ & $\f$ & $\af$ & 1 & $\cdots$ & 1 \\
  $Q_2$ & 1 & $\f$ & $\af$ & $\cdots$ & 1 \\
  $\vdots$ & $\vdots$ & $\vdots$ & $\vdots$ & $\ddots$ & $\vdots$ \\
  $Q_N$ & $\af$ & 1 & 1 & $\cdots$ & $\f$
\end{tabular}
\caption{The matter content}
\label{model}
\end{center}
\end{table}
It is shown that this simple model can imitate a five-dimensional
theory with bulk gauge multiplets~\cite{ACG,HPW}. Consider the link
variables $Q_i$ develop vacuum expectation values proportional to the
identity, 
$\langle Q_1\rangle=\cdots=\langle Q_N\rangle=v$.\footnote{The
diagonal form of VEVs is provided, for example, by a superpotential
introduced in~\cite{ACG,CEGK}, which gives (supersymmetry-breaking) 
masses only to the trace part of $Q_i$, and does not affect the 
discussion below.}  Below the scale $\sim gv$, the gauge symmetries
are reduced to a diagonal subgroup $G$ and the other gauge multiplets
become massive with discrete mass spectrum. This just looks like a 
five-dimensional $G$ gauge theory compactified on a circle $S^1$,
resulting Kaluza-Klein mass spectra. Note here that it is a simple
assumption for the bulk theory being five-dimensional Lorentz
invariant that the gauge couplings and VEVs of $Q_i$ take the common
values. This way of deconstructing or latticized dimensions is
useful in that one can study higher-dimensional theories from a
familiar four-dimensional point of view.

We here briefly review the mass spectrum of gauge bosons in this
model~\cite{ACG,HPW}. The complete Lagrangian and mass spectra are
given later. The mass matrix is derived from the
K\"ahler term of $Q_i$ fields, which gives 
\begin{eqnarray}
 {\cal L} &=& \frac{1}{2}k^2g^2v^2 A_\mu^iM_{ij} A^{\mu\,j},
 \label{mass}
\end{eqnarray}
where we have not written the implicit dependence of gauge indices,
and $k$ is the normalization factor of link variables that could
depend on the gauge coupling $g$ (see the Lagrangian (\ref{lag})). The
matrix $M$ is  
\begin{eqnarray}
  M &=& \pmatrix{
    2 & -1 & & & -1 \cr
    -1 & 2 & -1 & & \cr
    & \ddots & \ddots & \ddots & \cr
    & & -1 & 2 & -1 \cr
    -1 & & & -1 & 2}.
  \label{M}
\end{eqnarray}
From this, one obtains the mass eigenvalues $m_n^2$ and the
corresponding eigenstates $\widetilde A_n$, labelled by an integer $n$
(the Kaluza-Klein level),
\begin{eqnarray}
  m_n^2 &=& 4k^2g^2v^2\sin^2 \frac{n\pi}{N}, \label{KKmass} \\
  \widetilde A_\mu^n &=& \frac{1}{\sqrt{N}} \sum_{j=1}^N
  (\omega_n)^j A_\mu^j, \qquad\quad (n=0,\cdots,N-1)
\end{eqnarray}
where $\omega_n=e^{2\pi in/N}$. One can see that there is a massless
gauge boson and in addition the Kaluza-Klein tower of massive gauge
fields, the low-lying modes ($n\ll N$) of which gauge bosons have
masses approximately written as
\begin{eqnarray}
  m_n &\simeq& 2kgv\frac{n\pi}{N} \;=\; \frac{n}{R},
\end{eqnarray}
where we identify the compactification radius as $2\pi R=N/kgv$. The
mass term (\ref{mass}) becomes the kinetic energy transverse to the
four-dimensions in the continuum limit ($N\to\infty$).

\section{Moduli and supersymmetry-breaking scenarios}
\setcounter{equation}{0}

\subsection{Moduli}

A supersymmetry-breaking scenario in this type of models was examined
in~\cite{CEGK,CKSS} assuming that supersymmetry-breaking dynamics is
on one endpoint of the lattice sites. From a five-dimensional
viewpoint, that corresponds to supersymmetry being broken only on a
four-dimensional space like the gaugino mediation~\cite{gaugino}.

In this work, we study supersymmetry breaking in the above
four-dimensional model. To have insights into bulk symmetry breaking,
there need to be some modulus fields which are commonly coupled to any
multiplet in the theory. Here we consider two candidates of these
moduli. One is the dilaton field $S$. One may define a modulus $S_i$
for each gauge group whose scalar component gives a gauge coupling
constant $g_i$. As noted before, however, they have to interact with a
universal strength in order for this model to describe a proper
five-dimensional theory (on the flat background). In what follows, we
therefore assume $S\equiv S_1=\cdots=S_N$. We have also assumed the
universal value $v$ for the VEVs of link variables. Another modulus we
will consider is referred to as $Q$ which gives this universal
VEV\@. The modulus $Q$ may be a normalized composite field of
$Q_i$'s. We take the modulus forms which are invariant under a
translation transverse to the four dimensions, for
simplicity. Non-universal values of couplings and VEVs may be
interpreted as the presence of brane-like interactions and/or curved
backgrounds, and that issue will be studied elsewhere.

These modulus fields may have some connections with spacetime
symmetries since the modulus $S$ corresponds to dilatation and $Q$
relates to a size of compactification radius. It should be noticed
that, exactly speaking, $S$ is neither four- nor five-dimensional
dilaton, and $Q$ might not be the radion correctly. In our model, all
of these are not independent variables as seen below.

Let us discuss the relations between these combinations of modulus
fields. First we have the dilaton $S$ and the modulus $Q$ whose
VEVs are assumed to be
\begin{eqnarray}
  S &=& \frac{1}{4g^2} + F_S \theta^2, \label{S} \\
  Q &=& v + F_Q \theta^2. \label{Q}
\end{eqnarray}
In addition to these, we define the (combinations of) moduli fields
that give the following VEVs; 
\begin{eqnarray}
  S_4 &=& \frac{1}{4g_4^2} + F_{S4} \theta^2, \\
  S_5 &=& \frac{1}{4g_5^2} + F_{S5} \theta^2, \\
  T &=& \frac{1}{R} + F_T \theta^2,
\end{eqnarray}
where $g_4$, $g_5$ are the effective four- and five-dimensional gauge
couplings, and $R$ is the compactification radius of extra
dimensions. By comparing the low-energy description of the model (at
the energy below $\sim v$) with Kaluza-Klein theory, the following
tree-level relations among the parameters are
found~\cite{ACG,HPW}\footnote{The 1PI and holomorphic gauge couplings
differ only at higher-loop level in perturbation theory.},
\begin{eqnarray}
  \frac{1}{2\pi R} \;=\; \frac{kgv}{N}, \qquad
  g_4^2 \;=\; \frac{g^2}{N}.
  \label{relation}
\end{eqnarray}
The first equation is required to match the spectrum to that of
Kaluza-Klein theory, and the second equation can be regarded as a
volume suppression of bulk gauge coupling. In addition, the
five-dimensional gauge coupling is defined as (irrespectively of how
to get a five-dimensional model)
\begin{eqnarray}
  g_5^2 &=& 2\pi Rg_4^2,
  \label{5T4}
\end{eqnarray}
which comes from the normalization of gauge kinetic terms. These
relations among the couplings suggest that the modulus fields satisfy
the relations 
\begin{eqnarray}
  S_4 &=& N S, \\[1mm]
  S_5 &=& \frac{1}{2}QS^{1/2}k(S), \\
  T &=& \frac{\pi}{N}QS^{-1/2}k(S).
  \label{Trel}
\end{eqnarray}
The appropriate form of the factor $k(S)$ will be fixed in the next
section by holomorphy and other phenomenological arguments. From
these, we see that $S_4$, $S_5$ and $T$ are expressed in terms of two
moduli $S$ and $Q$. Of course every choice of two independent moduli
such as $(S,S_5)$, $(S_4,T)$, etc., can describe the same physics, and
in the present four-dimensional model, a natural choice is
$(S,Q)$. Each set of non-vanishing $F$-terms corresponds to one
supersymmetry-breaking scenario.

Extracting the $\theta^2$ terms, we obtain the $F$-components of
moduli
\begin{eqnarray}
  F_{S4} &=& NF_S, \\[1mm]
  F_{S5} &=& \frac{kv}{4g}\left(\frac{F_Q}{v}+2g^2
    \biggl(1+2\biggl\langle\frac{\partial\ln k(S)}{\partial\ln S}
    \biggr\rangle\biggr) F_S\right), \\ 
  F_T &=& \frac{2\pi kgv}{N} \left(\frac{F_Q}{v}-2g^2
    \biggl(1-2\biggl\langle\frac{\partial\ln k(S)}{\partial\ln S}
    \biggr\rangle\biggr) F_S\right).
\end{eqnarray}
It is emphasized that the four-dimensional dilaton $S_4$ is almost
close to the dilaton $S$, but its $F$-term satisfies the relation 
\begin{eqnarray}
  \frac{F_{S4}}{\langle S_4\rangle} &=& 
  \frac{F_{S5}}{\langle S_5\rangle} -\frac{F_T}{\langle T\rangle},
  \label{5T4-F}
\end{eqnarray}
independently of the detailed form of $k(S)$. Notice that this relation
comes out through the equation (\ref{5T4}), which implies $S_4$
depends on the radion $T$ and the five-dimensional dilaton $S_5$.

In the next section, we will discuss supersymmetry-breaking effects of
these moduli and calculate sparticle mass spectrum of the model. When
introducing appropriate potentials for the modulus fields, their VEVs
are fixed to some region or point in the moduli space of vacua. For
example, since $S$ is the dilaton for each gauge group, dilaton
stabilization mechanisms proposed in the literature are easily
incorporated in our framework. The situation is similar for the
modulus $Q$, corresponding to the radion. Moreover in describing
five-dimensional theory, $Q$ is actually assumed to be stabilized by
relevant superpotential terms as in Refs.~\cite{ACG,CEGK}. It is
therefore understood that deformation of (superpotential) terms could
also induce a supersymmetry-breaking VEV of $Q$. \ However, instead of
doing that, we study a more generic case. That is, in this letter we
investigate the whole parameter space of the moduli $F$-terms, and
then focus on several limits corresponding to bulk
supersymmetry-breaking scenarios. We do not try to construct specific
dynamics for modulus fields to have five-dimensional nature by tuning
potential couplings, since our aim here is not to present
five-dimensional theories. It is only the specific region of moduli
space where our model reproduces the known bulk supersymmetry-breaking
scenarios. In other words, the present framework contains unexplored
four-dimensional phenomena of supersymmetry breaking. It should be
noted that the tree-level mass formulae given in the next section are
not modified by the existence of moduli potentials. The only possible
case where the mass formula might be affected is that potentials for
moduli stabilization contain the multiplets for which one wants to
calculate their spectrum. We do not consider such a peculiar case in
this letter.

\subsection{Supersymmetry breaking in the bulk}

So far various supersymmetry breaking models have been discussed in
the literature within the frameworks of being concerned with higher
dimensional physics, and several examples are mentioned in the
Introduction. In the following, we will particularly focus on the
dilaton and moduli dominated supersymmetry breaking in the
string-inspired four-dimensional supergravity~\cite{sugra} and 
supersymmetry breaking by the radion $F$-term~\cite{noscale,radion}. 
Here one should pay attention to relevant choices of modulus $F$-terms
in examining supersymmetry-breaking models. That is, four-dimensional
(low-energy effective) theories know $F_{S4}$ and $F_T$ as fundamental
quantities, but on the other hand, five-dimensional ones $F_{S5}$ and
$F_T$. This point is important to the following discussion.

The dilaton dominance scenario is the four-dimensional supergravity
specified by a non-vanishing $F$-term of the four-dimensional dilaton
$S_4$ and negligible contribution from the overall modulus $T$. We
find from the result in the previous section that in the model where
the appropriate modulus fields are $S$ and $Q$, the scenario is
described by $F_S\neq0$ and $F_Q=2g^2v(1-
2\langle\frac{\partial\ln k(S)}{\partial\ln S}\rangle) F_S$. The VEVs
of the four and five-dimensional dilaton $F$-terms are then found to
be $F_{S4}/\langle S_4\rangle=F_{S5}/\langle S_5\rangle=
F_S/\langle S\rangle$. 

On the other hand, the moduli domination is also the four-dimensional
model characterized by the opposite limit of $F$-terms; a non-zero
$F_T$ and a vanishing dilaton $F$-term, $F_{S4}=0$. As a typical
spectrum of this scenario, gauginos are massless at tree level. This
is because the string perturbation theory shows that the gauge kinetic
function, which induces gaugino masses, depends only on $S_4$ at tree
level. This limit of $F$-terms is translated into the present model as
$F_S=0$ and $F_Q\neq0$. The other modulus $F$-components are then
given by $F_{S5}=(k/4g)F_Q$ and $F_T=(2\pi kg/N)F_Q$. 

The field-theoretical model similar to the moduli dominated
supersymmetry breaking is discussed in~\cite{KK}. This Kaluza-Klein
mediation model is a four-dimensional effective theory and has
the identical $F$-term VEVs with those in the moduli domination.
Sparticle mass spectra in this case are easily calculated from
renormalization-group functions in four dimensions, and the mechanism
has a wide variety of realistic model construction.

A related idea utilizing $F_T$ supersymmetry breaking is suggested in
the radion mediation model. It is a five (or higher) dimensional
model, and therefore a reasonable choice of two independent moduli is
$T$ and $S_5$. The radion mediation is thus defined by $F_T\neq0$ and
$F_{S5}=0$. In turn, this corresponds to $F_S\neq0$ and $F_Q=-2g^2v(1+
2\langle\frac{\partial\ln k(S)}{\partial\ln S}\rangle) F_S$ in the
present model. As a result, the four-dimensional dilaton $F$-term
becomes 
\begin{eqnarray}
  F_{S4} &=& \frac{-N}{2g^2v}\biggl(1
  +2\biggl\langle\frac{\partial\ln k(S)}{\partial\ln S}\biggr\rangle
  \biggr)^{-1}F_Q \;=\; \frac{-N^2}{8\pi kg^3v}F_T.
\end{eqnarray}
This means the four-dimensional gaugino 
mass $m_\lambda=-F_{S4}/2\langle S_4\rangle=F_T/2\langle T\rangle$,
which agrees with the result of zero-mode gaugino mass in~\cite{radion}.

In this way, we show via deconstruction that various known
supersymmetry-breaking scenarios can be seen by the difference in the
choices of non-zero modulus $F$-terms (as summarized in
Table~\ref{Fterm}). The parameter space spanned by two independent
$F$-terms is therefore the space of supersymmetry breaking in the
bulk, and several special limits in this parameter space correspond to
the scenarios which have been discussed in the literature.
\begin{table}[htbp]
{\small
\begin{center}
\renewcommand{\arraystretch}{1.8}
\begin{tabular}{|l||c|c||c|c|c|} \hline
  & $F_S$ & $F_Q$ & $F_{S4}$ & $F_{S5}$ & $F_T$ \\ \hline
  \parbox{18mm}{{\normalsize dilaton}\\ ($F_T=0$)} &
  $F_S$ & $2g^2v(1-x)F_S$ & $NF_S$ & $kgvF_S$ & 0 \\ \hline
  \parbox{18mm}{{\normalsize moduli}\\ ($F_{S4}=0$)} & 
  0 & $F_Q$ & 0 & $\displaystyle{\frac{k}{4g}F_Q}$ & 
  $\displaystyle{\frac{2\pi kg}{N}F_Q}$ \\ \hline
  \parbox{18mm}{{\normalsize radion}\\ ($F_{S5}=0$)} &
  $\displaystyle{\frac{-F_Q}{2g^2v(1+x)}}$ & $F_Q$ & 
  $\displaystyle{\frac{-NF_Q}{2g^2v(1+x)}}$ & 0 & 
  $\displaystyle{\frac{4\pi kg}{N(1+x)}F_Q}$ \\ \hline
\end{tabular}
\caption{{\normalsize The moduli $F$-terms and the typical
supersymmetry-breaking models in the bulk. The parameter $x$ is 
defined by $x\equiv 
2\langle\frac{\partial\ln k(S)}{\partial\ln S}\rangle$. The
holomorphy and some phenomenological arguments suggest $x=1$ and
$k=1/g$.}}
\label{Fterm}
\end{center}}
\end{table}

\section{Spectrum}
\setcounter{equation}{0}

In this section, we explicitly show the resulting
supersymmetry-breaking spectrum of Kaluza-Klein modes. We here focus
on the vector multiplets, but the quantitative aspects discussed below
are completely the same for bulk hypermultiplets.

Since we consider broken gauge symmetries and massive gauge fields, it
is convenient to use the unitary gauge for vector multiplets. In this
gauge, the goldstone chiral multiplets (the fluctuations around the
VEVs (\ref{Q})) are absorbed into the vector multiplets with suitable
gauge transformations. Consequently each vector multiplet contains a
massive vector field and two spinors, gaugino and goldstone
fermion. In addition, other dynamical and auxiliary bosonic components
are introduced. The link variables $Q_i$'s are then treated as
background fields with non-zero VEVs.

First it is easily found that the gauge fields do not get
supersymmetry-breaking contribution, and the mass spectrum is just
given by that calculated in Section 2; one massless gauge multiplet
corresponding to the diagonal subgroup $G$ and the Kaluza-Klein tower
with discrete mass spectrum (\ref{KKmass}).

The gauge fermion masses with supersymmetry breaking are calculated as
follows. The relevant piece of Lagrangian is
\begin{eqnarray}
  {\cal L} &=& \sum_i\left[\int d^2\theta\, SW_iW_i +\textrm{h.c.}\;
    +\int d^2\theta d^2\bar\theta\,K(S,S^\dagger)\,
    Q_i^\dagger e^{\sum V} Q_i\right].
  \label{lag}
\end{eqnarray}
We have included the universal couplings of the dilaton $S$. The
relevant field to appear here is $S$, and not the effective four or
five-dimensional dilaton $S_4$, $S_5$. The real 
function $K(S,S^\dagger)$ fixes the overall scale of discrete mass
spectra of this model ($k=\langle K|_{\theta=0}\rangle^{1/2}$) and its 
form will be determined later. Inserting the VEVs of (\ref{S}) and
(\ref{Q}), the mass terms take the following form;
\begin{equation}
  {\cal L}_\textrm{\footnotesize mass} \;=\; 
  -F_S \lambda_i\lambda_i -k^2v^2\chi_iM_{ij}\lambda_j 
  +\frac{1}{2}k^2v\Bigl(F_Q +v\Bigl\langle{
    \textrm{$\frac{\partial\ln K(S,S^\dagger)}{\partial\ln S}$}}
  \Bigr\rangle F_S\Bigr) \chi_iM_{ij}\chi_j \,+\textrm{h.c.}\,,
\end{equation}
where $\chi_i$ is the goldstone fermion now included in the vector
multiplet $V_i$. The first term comes from the gauge kinetic term
and the last two are induced by the tree-level K\"ahler term of $Q_i$,
so the flavor structure is the same as that of the gauge fields, which
is explained by the matrix $M$~(\ref{M}). Since $M$ also defines the
kinetic terms of $\chi_i$'s, the canonically normalized fields are
obtained by the redefinition $kvP\chi\to\chi$ where $P$ is a
square-root of $M$ ($M=P^{\rm t} P$) and written by
\begin{eqnarray}
  P \;=\; \pmatrix{
    1 & -1 & & \cr
    & \ddots & \ddots & \cr
    & & 1 & -1 \cr
    -1 & & & 1 }.
\end{eqnarray}
With this redefinition and a rescaling $\lambda\to g\lambda$, the mass
matrix of the normalized spinors becomes
\begin{eqnarray}
  {\cal L}_\textrm{\footnotesize mass} &=&
  -\frac{1}{2}\pmatrix{\lambda & \chi} 
  \left(\begin{array}{cc}
      2g^2F_S & kgvP^{\rm t} \\[1mm]
      kgvP & -\frac{F_Q}{v}
      -\Bigl\langle\frac{\partial\ln K(S,S^\dagger)}{\partial S}
      \Bigr\rangle F_S
    \end{array} \right)
  \pmatrix{\lambda \cr \chi} +\textrm{h.c.}\,.
\end{eqnarray}

Without the $F$-term contributions, the mass eigenstates take the same
form as the gauge fields. This is an indication of $N=2$
supersymmetry, equivalently $N=1$ supersymmetry in five dimensions,
which is expected to appear in the infrared. In this mass basis of
$\widetilde\lambda_n$ and $\widetilde\chi_n$, the mass matrix is
rewritten as follows
\begin{eqnarray}
  {\cal L}_\textrm{\footnotesize mass} &=& -\frac{1}{2}
  \pmatrix{\widetilde\lambda & \widetilde\chi}
  \left(\begin{array}{cc}
      2g^2F_S & B \\[1mm] B & -\frac{F_Q}{v}
      -\Bigl\langle\frac{\partial\ln K(S,S^\dagger)}{\partial S}
      \Bigr\rangle F_S
    \end{array} \right)
  \pmatrix{\widetilde\lambda \cr \widetilde\chi} +\textrm{h.c.}\,,
  \label{matrix}
\end{eqnarray}
where the elements of the diagonal matrix
$B_{ij}=2kgv\sin\frac{j\pi}{N}\;\delta_{ij}$ are the Kaluza-Klein
Dirac masses. The irrelevant phase factors have been absorbed with
field redefinitions. We finally 
obtain the mass eigenvalue of the level--$n$ Kaluza-Klein gauge
fermions ($n=0,\cdots,N-1$);
\begin{eqnarray}
  m_{\lambda_n} &=& \frac{1}{2}\left[
    \pm\sqrt{4m_n^2+\biggl(\frac{F_Q}{v}+2g^2
      \biggl(1+2\Bigl\langle{\textrm{\large 
        $\frac{\partial\ln K(S,S^\dagger)}{\partial\ln S}$}}
      \Bigr\rangle\biggr) F_S\biggr)^2\,} \right. \nonumber \\[1mm]
  && \Biggl. \hspace{2cm} -\frac{F_Q}{v}+2g^2
    \biggl(1-2\Bigl\langle\textrm{\large 
      $\frac{\partial\ln K(S,S^\dagger)}{\partial\ln S}$}
    \Bigr\rangle\biggr) F_S \;\Biggr],
  \label{KKgaumass}
\end{eqnarray}
where $m_n$ is the Kaluza-Klein mass of gauge fields (\ref{KKmass}),
which is supersymmetric contribution. The positive (negative)
sign in the bracket corresponds to the gaugino (the goldstone fermion)
mass. Here the states which are equal 
to $\widetilde\lambda$, $\widetilde\chi$ in the supersymmetric limit
are referred to as gauginos and goldstone fermions, respectively. It
is interesting to note that the gauge fermion mass (\ref{KKgaumass})
can be more simply expressed with only the five-dimensional
quantities:
\begin{eqnarray}
  m_{\lambda_n} &=& \frac{1}{2}\left[ \pm\sqrt{4m_n^2
      +\biggl(\frac{F_{S5}}{\langle S_5\rangle}\biggr)^2\,}
    -\frac{F_T}{\langle T\rangle}\,\right].
  \label{KKgaumass2}
\end{eqnarray}
This result implies that higher-dimensional effects, even including
supersymmetry breaking, are properly reproduced in our model.

We now examine our result for the supersymmetry-breaking models
discussed in the previous section.

\smallskip
\noindent $\bullet$ Dilaton dominated supersymmetry breaking

This scenario is characterized by the limit $F_T=0$. We then obtain
the Kaluza-Klein masses with the supersymmetry-breaking effect
\begin{eqnarray}
  m_{\lambda_n}\textrm{\small (dilaton)} &=& 
  \pm\sqrt{m_n^2+(2g^2F_S)^2 }.
\end{eqnarray}
The spectrum is just as expected in the dilaton dominant case in
supergravity models. The first term in the square-root is the
Kaluza-Klein Dirac mass, and the second one is a
supersymmetry-breaking part that is certainly provided by the
four-dimensional dilaton coupling 
($2g^2F_S=F_{S4}/2\langle S_4\rangle$). Note that all the 
Kaluza-Klein states including zero modes receive the universal
supersymmetry-breaking contribution. The two level--$n$ spinors are
degenerate in mass, and the mass splitting between bosons and fermions
are equal for all Kaluza-Klein modes. This fact is regarded as a
reflection that the dilaton field (the action of dilatation) commonly
couples to any field in the theory. The universal mass spectrum is one
of the major motivations to investigate the dilaton dominant limit in
supergravity models. The universality in our model is more clearly
seen for scalar components in hypermultiplets. In that case, taking
the $F_T=0$ limit washes away the bulk mass dependence of
supersymmetry-breaking scalar masses~\cite{next}.

\medskip
\noindent $\bullet$ Moduli dominated supersymmetry breaking 

With the definition of $F_{S4}=0$, the gauge fermion mass spectrum
becomes
\begin{eqnarray}
  m_{\lambda_n}\textrm{\small (moduli, {\footnotesize KK})} &=& 
  \pm\sqrt{m_n^2+\left(\frac{F_Q}{2v}\right)^2} -\frac{F_Q}{2v}.
  \label{modu-KKgau}
\end{eqnarray}
It is interesting that even when supersymmetry breaking is turned on,
the zero-mode gaugino is massless and does not get a mass splitting
with the zero-mode gauge field. (The $n=0$ spinor being affected by
the non-zero $F$-terms is the goldstone fermion $\widetilde\chi_0$.)
This is exactly the spectrum predicted in this type of
supersymmetry-breaking scenario~\cite{sugra,KK}. By definition, the
scenario assumes a vanishing $F$-term of the four-dimensional
dilaton. The zero-mode gaugino mass is then shifted at loop level by
string threshold corrections or effects of bulk fields. In our model,
the spectrum is easily read from the mass matrix (\ref{matrix}). The
gaugino $\widetilde\lambda_0$ is massless due to the vanishing $F_S$
and Kaluza-Klein mixing mass. As for the excited modes, the
supersymmetry-breaking contribution from $F_Q$ is transmitted to
gauginos through the non-zero Kaluza-Klein masses. The situation is
similar to the models where gauge multiplets behave as messengers, and
sparticle soft masses at loop level are calculated from wave-function
renormalization in four dimensions~\cite{GR}. Therefore our approach
is also likely to describe this limit well.

There may be an intuitive explanation for this type of spectrum as was
discussed in Ref.~\cite{KK}. That is, a non-zero $F$-term of the
modulus which gives Kaluza-Klein masses does not induce tree-level
supersymmetry-breaking masses for zero modes, as these two mass terms
are proportional to Kaluza-Klein numbers. In the present case, such a
modulus 
corresponds to the one whose scalar component obtains a VEV 
$\propto 1/R$, and is given by $T\propto Q$. This interpretation
becomes manifest in examining mass spectra of bulk hypermultiplets
with moduli fields~\cite{next}.

\medskip
\noindent $\bullet$ Radion $F$-term breaking 

This scenario takes the $F$-term assumption $F_{S5}=0$, that is
converted into $F_Q=-2g^2v(1+
2\langle\frac{\partial\ln k(S)}{\partial\ln S}\rangle) F_S$. We find
that the gaugino mass matrix (\ref{matrix}) in this 
limit has the exactly same form as calculated in Ref.~\cite{MP}, where
they use an $N=1$ superfield formalism of the five-dimensional action
with the radion superfield. The mass eigenvalues of the Kaluza-Klein
spinors become 
\begin{eqnarray}
  m_{\lambda_n}\textrm{\small (radion)} &=& \pm m_n-\frac{F_Q}{v}
  -\biggl\langle\frac{\partial\ln k(S)}{\partial S}\biggr\rangle F_S.
  \label{rad-KKgau} 
\end{eqnarray}
The scenario assumes a non-zero value of the radion $F$-term. However,
compared to the moduli dominance scenario stated above, there is a
difference in a contribution from the dilaton field $S$, resulting the
non-zero $F$-term of the four-dimensional dilaton $S_4$. This gives a
tree-level mass of the gaugino zero mode. In other words, if the
moduli domination were seen from a five-dimensional viewpoint, there
would appear to be an additional contribution from $S_5$ such that the
definition $F_{S4}=0$ is preserved (see Eq.~(\ref{5T4-F})). On the
other hand, the masses of the Kaluza-Klein excited modes are rather
similar to each other. In particular, the low-lying modes have masses 
\begin{eqnarray}
  m_{\lambda_n}\textrm{\small (moduli, {\footnotesize KK})} \;=\;
  m_{\lambda_n}\textrm{\small (radion)} \;\simeq\; 
  \pm\frac{n}{R}-\frac{R}{2}F_T,
\end{eqnarray}
where we have assumed that the supersymmetry-breaking part is smaller
than the supersymmetric Kaluza-Klein mass (i.e., $RF_T\ll v$).

It has been shown~\cite{MP,GQ} that the radion mediation model has
the same spectrum as that predicted by the Scherk-Schwarz
mechanism~\cite{SS}. The Scherk-Schwarz theory is essentially higher
dimensional and adopts twisted boundary conditions for bulk fields
along the extra dimensions. On the other hand, the moduli dominated
supersymmetry breaking in four-dimensional supergravity (and the
Kaluza-Klein mediation) is not a Scherk-Schwarz theory and does gives
different soft terms, as explicitly shown in the above.

\bigskip

Let us finally discuss the normalization function $K(S,S^\dagger)$ in
the Lagrangian (\ref{lag}). It should be mentioned that the form of
gaugino masses (\ref{KKgaumass2}) is not affected by any details of
the factor $K(S,S^\dagger)$, and the above qualitative discussions
about the gaugino mass spectrum are generic and still preserved.
We propose the proper form of $K$ is given by
\begin{eqnarray}
  K(S,S^\dagger) &=& \frac{8}{1/S+1/S^\dagger}.
  \label{K}
\end{eqnarray}
The factors $k$ and $k(S)$ defined in Section 2 then become $k=1/g$ and
$k(S)=2S^{1/2}$, respectively. Though the complete form of $K$ is not
determined without referring to higher-dimensional physics, (\ref{K})
is found to be certainly consistent with several non-trivial and
independent requirements. First, notice that to have right results
based on holomorphy, the normalization of the link variables $Q_i$'s
is required to be $\langle K\rangle=1/g^2$. With this choice, the
gauge and adjoint chiral multiplets of the low-energy $G$ gauge theory
have the same field normalization. Moreover, in this case, the radion
superfield in our model becomes independent of the dilaton superfield
(see the relation (\ref{Trel})), which result is plausible since, for
example, it does not lead to an undesirable relation between the theta
angle and the graviphoton field.

Secondly, with an explicit form of $K(S,S^\dagger)$, one can evaluate
tree-level masses of the scalar fields of $Q_i$'s. They are the
adjoint scalar fields of the low-energy $G$ gauge theory, and are
contained in vector multiplets of the enhanced $N=2$
supersymmetry. The scalar mass $m_c^2$ generated by the K\"ahler term
with (\ref{K}) is
\begin{eqnarray}
  m_{c_n}^2 &=& m_n^2 +2\,\textrm{Re}
  \left(\frac{F_{S5}^{\,*}}{\langle S_5\rangle}
    \frac{F_T}{\langle T\rangle}\right).
\end{eqnarray}
Let us examine this mass formula in the limits discussed before. One
can easily see that the radion mediation limit ($F_{S5}=0$) does not
give supersymmetry-breaking soft mass. This indeed agrees with the
fact that the radion mediation is equivalent to the Scherk-Schwarz
mechanism, which is now applied to the $SU(2)_R$ symmetry under which
the adjoint scalars are singlet and hence do not get symmetry-breaking
masses. If one first requires that the scalars $c_n$ do not have soft
terms in the $F_{S5}=0$ limit, $K(S,S^\dagger)$ has to satisfy 
$\bigl\langle\frac{\partial\ln K}{\partial S\partial S^\dagger}
\bigr\rangle=-(2g^2)^2$. The most probable solution of this equation
is $K=X(S)X(S^\dagger)/(S+S^\dagger)$, where $X$ is an arbitrary
function. Then the holomorphy argument suggests $X(S)\propto S$ and
thus (\ref{K}). For completeness, we write down the scalar masses in
the other limits;
\begin{eqnarray}
  m_{c_n}^2\textrm{\small (dilaton)} &=& m_n^2, \qquad
  m_{c_n}^2\textrm{\small (moduli, {\footnotesize KK})} \;=\; m_n^2
  +2\left|\frac{F_T}{\langle T\rangle}\right|^2.
\end{eqnarray}

The third consistency is about the 5-5 component of the
five-dimensional metric, $g_{55}$. In a continuum five-dimensional
theory, the kinetic energy terms of bosonic fields along the fifth
dimension have a dependence on $g_{55}$ as
$\sqrt{g_{55}}\,g^{55}\,\propto\,1/R$. In the model at hand, the second
term in the Lagrangian (\ref{lag}) becomes this kinetic energy in the
continuum limit, and its modulus dependence is given by 
$\langle K(S,S^\dagger)\,Q^\dagger Q\rangle$. The equation (\ref{K})
then indicates $\langle KQ^2\rangle$ $\sim$ $\langle SQ^2\rangle$ $\sim$
$\langle S_5T\rangle$. As a result, the desirable metric dependence
appears, for a fixed value of the five-dimensional gauge coupling
$g_5$. 

\bigskip

We close this section by a comment on the model which turns into a
five-dimensional theory compactified on an $S^1/Z_2$ orbifold. This
can be formulated~\cite{HPW,CEGK} by getting rid of a link variable,
e.g.~$Q_N$, from the $S^1$ model. In this case, additional fields may
be introduced to cancel gauge anomalies on the orbifold fixed
points. Examining a mass matrix, it is found that $Q_i$'s contain only
massive modes, and the zero-mode state consists of an $N=1$ vector
multiplet without an associated adjoint chiral multiplet, which
situation corresponds to the $Z_2$ orbifolding. In turn, this results
in removing $\widetilde\chi^0$ and $c_0$ in our analyses. The plus
sign is chosen for the zero mode, and the gaugino masses in various
limits discussed before are not altered. Results similar to those in
the $S^1$ case hold for other quantities, for example, the
Kaluza-Klein mass spectrum is unchanged except for a 
replacing $N\to 2N$ $(R\to 2R)$.

\section{Summary}
\setcounter{equation}{0}

In this paper, we have studied supersymmetry breaking in the
four-dimensional model with two types of modulus fields. The model
can describe five-dimensional physics in the infrared, and given the
relations among the modulus fields, we have discussed supersymmetry
breaking in the higher-dimensional bulk. The analysis is based on a
four-dimensional model, that is renormalizable and calculable in a
usual manner. We have made it clear that several specific limits in
the two-dimensional parameter space of the modulus $F$-terms
correspond to the bulk supersymmetry-breaking scenarios in the
literature. We have shown this by examining the gaugino and adjoint
scalar masses in the cases of the $S^1$ and $S^1/Z_2$
compactifications. It is non-trivial to establish such correspondences
and indeed require a properly-fixed moduli dependence of the
action. The moduli dependence will also be confirmed by detailed
calculations of radiative corrections to mass
spectrum~\cite{next}. Moreover it would be an interesting issue to
study other choices of couplings and limits, which could describe
unexplored supersymmetry breaking in four or higher dimensions, and we
leave it to future work. Besides the issue of supersymmetry breaking,
extra dimensions provides a new perspective for various subjects in
particle physics. Realistic model construction along this line of
using a purely four-dimensional one will deserve further
investigations.

\vspace*{5mm}
\subsection*{Acknowledgments}

The authors are grateful to H.~Nakano, H.~Terao for useful discussions
and comments, and to the Summer Institute 2001 held at Yamanashi,
Japan. N.M.~thanks D.E.~Kaplan for an interesting discussion. This
work is supported in part by the Japan Society for the Promotion of
Science under the Postdoctoral Research Program (No.~08557 and
No.~07864).



\newpage
\begin{thebibliography}{99}
\bibitem{GUT}
J.~Ellis, S.~Kelley and D.V.~Nanopoulos, {\sl Phys.~Lett.} {\bf 260B}
(1991) 131; U.~Amaldi, W.~de~Boer and H.~F\"urstenau, 
{\sl Phys.~Lett.} {\bf 260B} (1991) 447; P.~Langacker and M.~Luo,
{\sl Phys.~Rev.} {\bf D44} (1991) 817.
\bibitem{hierarchy}
E.~Witten, {\sl Phys.~Lett.} {\bf 105B} (1981) 267.
\bibitem{sugra} 
L.E.Ib\'a\~nez and D.~L\"ust, {\sl Nucl.~Phys.} {\bf B382} (1992) 305;
B.~de~Carlos, J.A.~Casas and C.~Mu\~noz, {\sl Phys.~Lett.} {\bf 299B}
(1993) 234; A.~Brignole, L.E.Ib\'a\~nez and C.~Mu\~noz, 
{\sl Nucl.~Phys.} {\bf B422} (1994) 125; T.~Kobayashi, D.~Suematsu,
K.~Yamada and Y.~Yamagishi, {\sl Phys.~Lett.} {\bf 348B} (1995) 402.
\bibitem{others}
I.~Antoniadis, S.~Dimopoulos, A.~Pomarol and M.~Quiros, 
{\sl Nucl.~Phys.} {\bf B544} (1999) 503; N.~Arkani-Hamed, L.~Hall,
D.~Smith and N.~Weiner, {\sl Phys.~Rev.} {\bf D63} (2001) 056003;
R.~Barbieri, L.J.~Hall and Y.~Nomura, {\sl Phys.~Rev.} {\bf D63}
(2001) 105007; T.~Gherghetta and A.~Riotto, {\sl Nucl.~Phys.} 
{\bf B623} (2002) 97.
\bibitem{ACG} 
N.~Arkani-Hamed, A.G.~Cohen and H.~Georgi, {\sl Phys.~Rev.~Lett.} 
{\bf 86} (2001) 4757.
\bibitem{HPW} 
C.T.~Hill, S.~Pokorski and J.~Wang, {\sl Phys.~Rev.} {\bf D64} (2001)
105005.
\bibitem{apply} 
H.~Cheng, C.T.~Hill, S.~Pokorski and J.~Wang, {\sl Phys.~Rev.} 
{\bf D64} (2001) 065007; N.~Arkani-Hamed, A.G.~Cohen and H.~Georgi, 
{\sl Phys.~Lett.} {\bf 513B} (2001) 232; {\sl JHEP} {\bf 0207} (2002)
020; hep-th/0108089; H.~Cheng, C.T.~Hill and J.~Wang, {\sl Phys.~Rev.} 
{\bf D64} (2001) 095003; K.~Sfetsos, {\sl Nucl.~Phys.} {\bf B612}
(2001) 191; C.~Csaki, G.D.~Kribs and J.~Terning, {\sl Phys.~Rev.} 
{\bf D65} (2002) 015004; H.~Cheng, K.T.~Matchev and J.~Wang, 
{\sl Phys.~Lett.} {\bf 521B} (2001) 308; P.H.~Chankowski, A.~Falkowski
and S.~Pokorski, {\sl JHEP} {\bf 0208} (2002) 003.
\bibitem{CEGK} 
C.~Csaki, J.~Erlich, C.~Grojean and G.~Kribs, {\sl Phys.~Rev.} 
{\bf D65} (2002) 015003.
\bibitem{CKSS}
H.C.~Cheng, D.E.~Kaplan, M.~Schmaltz and W.~Skiba, {\sl Phys.~Lett.} 
{\bf 515B} (2001) 395.
\bibitem{gaugino} 
D.E.~Kaplan, G.D.~Kribs and M.~Schmaltz, {\sl Phys.~Rev.} {\bf D62}
(2000) 035010; Z.~Chacko, M.A.~Luty, A.E.~Nelson and E.~Ponton, 
{\sl JHEP} {\bf 0001} (2000) 003.
\bibitem{noscale}
E.~Cremmer, S.~Ferrara, C.~Kounnas and D.V.~Nanopoulos, 
{\sl Phys.~Lett.} {\bf 133B} (1983) 61; J.R.~Ellis, A.B.~Lahanas,
D.V.~Nanopoulos, and K.~Tamvakis, {\sl Phys.~Lett.} {\bf 134B} (1984)
429; J.R.~Ellis, C.~Kounnas, D.V.~Nanopoulos, {\sl Nucl.~Phys.} 
{\bf B247} (1984) 373; A.B.~Lahanas and D.V.~Nanopoulos, 
{\sl Phys.~Rept.} {\bf 145} (1987) 1.
\bibitem{radion} 
Z.~Chacko and M.A.~Luty, {\sl JHEP} {\bf 0105} (2001) 067.
\bibitem{KK} 
T.~Kobayashi and K.~Yoshioka, {\sl Phys.~Rev.~Lett.} {\bf 85} (2000)
5527.
\bibitem{next} 
T.~Kobayashi, N.~Maru and K.~Yoshioka, in preparation.
\bibitem{GR}
G.F.~Giudice and R.~Rattazzi, {\sl Nucl.~Phys.} {\bf B511} (1998) 25.
\bibitem{MP} 
D.~Marti and A.~Pomarol, {\sl Phys.~Rev.} {\bf D64} (2001) 105025.
\bibitem{GQ}
G.~von Gersdorff and M.~Quiros, {\sl Phys.~Rev.} {\bf D65} (2002)
064016.
\bibitem{SS}
J.~Scherk and J.H.~Schwarz, {\sl Phys.~Lett.} {\bf 82B} (1979) 60; 
{\sl Nucl.~Phys.} {\bf B153} (1979) 61.
\end{thebibliography}
\end{document}